\documentclass[12pt]{article}
\usepackage[margin=1in]{geometry}  % Sets margin size
\usepackage{amsmath,amsthm,amssymb,amsfonts}

\usepackage{bbm} % for fonts

\usepackage[dvipsnames]{xcolor}   %for color

\usepackage[utf8]{inputenc}   % for Spanish and French accents

\usepackage{hyperref}       % for hyperlinks within the document

\usepackage{natbib}

\newcommand{\Prob}{\ensuremath{\mathbb{P}}}

\newtheorem{theorem}{Theorem}
\theoremstyle{remark}
\newtheorem*{remark}{Remark}

\allowdisplaybreaks

\begin{document}

\title{\textbf{Optimal Group Size in Microlending}}
\author{
    Philip Protter\thanks{Supported in part by NSF grant DMS-1612758} \\
    \small{Statistics Department}\\
    \small{Columbia University}\\
    \small{New York, NY, 10027}\\
    \small{ORCID 0000-0003-1344-0403}\\
    \small{pep2117@columbia.edu}
    \and
    Alejandra Quintos\thanks{Supported in part by the Fulbright-García Robles Program} \thanks{Corresponding author} \\
    \small{Statistics Department}\\
    \small{Columbia University}\\
    \small{New York, NY, 10027}\\
    \small{ORCID 0000-0003-3447-3255}\\
    \small{a.quintos@columbia.edu}
}

% \author{Philip Protter\thanks{
% Statistics Department, Columbia University, New York, NY, 10027. \qquad ORCID: 0000-0003-1344-0403}
% \thanks{Supported in part by NSF grant DMS-1612758}\and
% Alejandra Quintos\thanks{
% Statistics Department, Columbia University, New York, NY, 10027. \qquad ORCID: 0000-0003-3447-3255}
% \thanks{Supported in part by the Fulbright-García Robles Program}
% \thanks{Corresponding author at a.quintos@columbia.edu}}

\date{\today}
\maketitle

\section*{ABSTRACT}
Microlending, where a bank lends to a small group of people without credit histories, began with the Grameen Bank in Bangladesh, and is widely seen as the creation of Muhammad Yunus, who received the Nobel Peace Prize in recognition of his largely successful efforts. Since that time the modeling of microlending has received a fair amount of academic attention. One of the issues not yet addressed in full detail, however, is the issue of the size of the group. Some attention has nevertheless been paid using an experimental and game theory approach. We, instead, take a mathematical approach to the issue of an optimal group size, where the goal is to minimize the probability of default of the group. To do this, one has to create a model with interacting forces, and to make precise the hypotheses of the model. We show that the original choice of Muhammad Yunus, of a group size of five people, is, under the right, and, we believe, reasonable hypotheses, either close to optimal, or even at times exactly optimal, i.e., the optimal group size is indeed five people.

% This paper uses a mathematical model with appropriate assumptions, to model and determine the optimal group size for microfinance loans.  An optimal size is defined to be the one that maximizes the probability of no default of the group.
\pagebreak

\section{INTRODUCTION}
Microfinance analyzes the lending mechanisms for people without access to traditional credit systems because of their low income, lack of collateral, or credit history. There are many lending mechanisms studied in the literature, but we will focus on group lending, which was introduced by Muhammad Yunus of the Grameen bank in Bangledesh. He won the Nobel Peace Prize (2006) for his efforts. In this mechanism, a loan is made to all members of the group for a fixed period of time (often less than a year). 

One of the crucial components of group lending is known as contingent renewal, which is a penalty that eliminates or reduces access to future loans to all members in a group if any of them defaults. This is to say, the default of at least one individual provokes the default of the group. Note that by default of a member we mean that she stops paying her share of the loan. See the articles \citealt{DDKP}, \citealt{DienerMauk}, and \citealt{DS},  for a study of the consequences of implementing various penalties for default.

The group lending mechanism is thought to be useful because it induces: (i) peer selection of members in a group since they are better informed than the lender about other potential borrowers, (ii) peer pressure to help enforce payments by other members in the group, and (iii) peer monitoring between the group members, but  particularly by a leader, to ensure continued performance on the loan. The structure of having a group leader is known in the literature as having an intermediary. From the lender’s point of view, it involves a delegation of the task of monitoring the loan. These ideas have, of course, been well studied in the Economics literature, and we rely on the seminal work of Philip Bond (\citealt{Bond2004}). In his paper, Bond shows that the joint liability of intermediary borrowers arises naturally in models of a financial intermediary as a delegated moderator. Bond shows that intermediation with joint liability is Pareto superior to intermediation without joint liability. As such, Bond builds on prior work of Morduch (\citealt{Morduch1999}) and Krasa and Villamil (\citealt{KrasaVillamil}).

The existing academic literature primarily focuses on understanding why the group lending mechanism is successful in reducing defaults. Both static and dynamic models have been analyzed (see \citealt{Stiglitz1990}; \citealt{Varian1990}; \citealt{Conlin1999}; \citealt{Morduch1999}; \citealt{Chowdhury2005}, \citealt{Chowdhury2007}; \citealt{Tedeschi2006}).
A related and somewhat unexplored issue is to determine an optimal group size. We define optimal size as the one that maximizes the probability of no default of the group. While cultural and other non-economic factors influence default, this paper focuses solely on group size.

The problem of an optimal group size has been analyzed by economists in the past (see \citealt{Armendariz}; \citealt{Gine}; \citealt{Ahlin2015}, \citealt{Ahlin2017}). Most of them take an intuitive, experimental and, often verbal approach. In contrast, ours is more probabilistic in nature, and thereby quantifiable. A first approach to what we present here is the article of \citealt{ProtterJarrow}. We also note that the problem of group size has been tackled using tools from Game Theory (see \citealt{Rezaei}).

An outline for this paper is as follows. Section II presents the model and our main theorem, while Section III outlines an interpretation for it. Section IV shows an example and discusses its intuitiveness, and Section V concludes. For easiness of readability, the proofs are included in the appendices.

\section{THE MODEL}
{\bf Caveats:} We begin with a few caveats. Different scenarios of Microlending have been considered in the academic literature. For example, there is the issue of the size of the loan affecting the group in group lending, and in particular its size (with larger loans leading to somewhat larger groups, see \citealt{Rezaei}). In this paper, the loan size is fixed, and we will not consider its influence on the group size. Some researchers take as their point of view the maximization of the profits of the lending institution (see, e.g., \citealt{BS}). This runs counter to the spirit embraced in \citealt{Yunus}, and again, while a valid consideration, it is not our concern in this paper. We are concerned only with minimizing the possibility of default on the loan; admittedly this is related to maximizing the profits of the lender, along with the interest rates charged (see \citealt{ProtterJarrow}). Finally, we mention that we do not discuss the transaction costs of banks, and what the effect of group lending is on them. We also implicitly assume that the members of a given group form a fairly homogeneous collection of people (see \citealt{JD}, later echoed in \citealt{BS}.) This homogeneity assumption is reflected in our assumption of identical distributions, within a group of a given size, allowing the distributions to change with the group size.

\bigskip
\noindent We now introduce the notation for our model. Let $N_i$ be the event of no default of member $i$ in a group of size $k$ $\left(k\in\mathbb{Z}^+, \ k\geq2\right)$, $\mathcal{N}_k$ be the event that the group of size $k$ does not default and, $\varphi(k):=1-\Prob_k\left(N_1\right)$, i.e., the probability of default of member 1 in a group of size $k$. Recall that, as explained in the introduction, the group lending mechanism implies that if at least one member of the group defaults, then the whole group defaults. This is what we call \textit{default of a group}.

We make the following assumptions:
\begin{enumerate}
    \item For fixed size $k$, the group members are independent and identically distributed.
    \item The probability of no default of one person depends on the size of the group. We make this explicit by writing: $\Prob_k\left(N_1 \right)$.
    \item $\Prob_k(N_1) > 0$. Otherwise the problem is trivial as the members will default for sure.
\end{enumerate}

We are interested in finding an optimal group size, that is, finding the number of people $k^*$ that maximizes the probability of no default of the group. Using our assumptions, along with our definition of default of a group, this translates into maximizing:
\begin{equation}\label{eq:to.maximize}
    \Prob\left(\mathcal{N}_k\right)= \Prob \left( \bigcap_{i=1}^k N_i \right) = \left[ \Prob_k\left(N_1 \right) \right]^k = \left( 1 - \varphi(k) \right)^k
\end{equation}
For a moment, suppose that $\varphi(\cdot)$ is constant in $k$, hence as $\varphi(\cdot)<1$, $\left(1-\varphi(\cdot)\right)^k$ decreases as $k$ increases. So, in order to have a maxima in (\ref{eq:to.maximize}), it makes sense to require that $1-\varphi(k)$ increases with $k$, which means that $\varphi(k)$ needs to decrease with $k$. The question is then, at what speed? This motivates the following theorem.

\begin{theorem}
    Let $\varphi(x) = \frac{1}{f(x)}$, for all $ x\in\mathbb{R}^+$
    If:
     \begin{enumerate}
         \item $f(x) > 1 \ \text{for all } x \geq 2 $
         \item $f(x)\in \mathcal{C}^2$ 
         \item $f'(x)>0 \ \text{for all } x \geq 2$
         \item \label{itm:4cond} There exist $a, b\in\mathbb{R} \ (a<b, \ a\geq 2)$ such that either:
         \begin{enumerate}
            \item \label{itm:4i}  $f(a)-af'(a)=\frac{1}{2}$ and $f(b)-bf'(b)=1$ \par
            and
            \item \label{itm:4ii}  $f''(x) < 0 \ \text{for all } x\in(a,b)$
         \end{enumerate}
         or
         \begin{enumerate}
            \setcounter{enumii}{2}          % to start enumerating with number 3
             \item \label{itm:4iii} $f(a)-af'(a)=1$ and $f(b)-bf'(b)=\frac{1}{2}$ \par 
             and
            \item \label{itm:4iv} $f''(x) > 0 \ \text{for all } x\in(a,b)$
         \end{enumerate}
     \end{enumerate}
    Then $\left(1-\varphi(x)\right)^x$ has a unique maximizer $x^*$ in $(a,b)$. Moreover, if $a$, $b$ are unique, then $x^*$ is the unique maximizer.
\end{theorem}
\textit{Proof.}
    See Appendix \ref{Proof.Theorem}

\vspace{3mm}

\begin{remark}
The conditions (\ref{itm:4i} and (\ref{itm:4iii}, which may seem mysterious at first glance, are inspired by Taylor's Theorem, from calculus.
\end{remark}

One can argue that, in the previous theorem, we heavily used the continuity of $f(x)$ and the fact that $x\in\mathbb{R}^+$ and, as we are optimizing with respect to the number of people, we should have taken $k\in\mathbb{Z}$, but we can always round $x^*$ to the closest integer to get $k^*$.

\section{INTERPRETATION OF THE THEOREM} 
Recall formula (\ref{eq:to.maximize}) $\Prob\left(\mathcal{N}_k\right)= \Prob \left( \bigcap_{i=1}^k N_i \right) = \left[ \Prob_k\left(N_1 \right) \right]^k = \left( 1 - \varphi(k) \right)^k$. As we briefly discussed in section (II), because of our independence and identical distribution assumptions, there are 2 interacting forces affecting $\Prob(\mathcal{N}_k)$. On the one hand, $\Prob(\mathcal{N}_k)= \left( 1-\varphi\left( k \right) \right)^k$ decreases as $k$ increases because $0<\varphi(k) <1$. On the other hand, we set \textit{a fortiori} $\varphi(k)$ to decrease as $k$ increases with the hope to find a maximizer $k^*$. Lending to a group has advantages over lending to an individual, but as the size of the group increases, the advantages diminish and tend to zero. There should, therefore, be some happy (and optimal) compromise of a group size being big, but not too big!

There are two opposing forces here. On the one hand, as the group size increases, the responsibility for performing one's tasks becomes dispersed, increasing the likelihood that one or more member of the group may default. Typically, there will be a leader or primary organizer, the force behind the loan, and she will need to ride herd on the other members, keeping them in line, if need be. The larger the group, the more diffused her efforts will be, and therefore the less effective.  Mathematically, as there are more people in the group, because of our assumptions, there are more independent chances of failure as it is riskier to have $k+1$ possible defaults than $k$. This causes $\Prob(\mathcal{N}_k)$ to decrease as $k$ increases. Note that the effect of this force is free of the choice of $\varphi(k)$.

On the other hand, as the group size increases, there are more collective resources (material and non material) which decrease the likelihood of default. In our model, this is captured by $\varphi(k)$ as different functions correspond to different contribution of the resources and hence different $k^*$ (optimal group size). Moreover, as the group size contracts, each person becomes more important and there are less resources available, making it harder to recover from a mistake, or a temporary period of misfortune. In the limit case of only one borrower, lenders in Ghana (for example) have found that, there being no peer pressure at all, the borrower has a serious probability of simply absconding with the money.~\footnote{Personal conversation of the first author in Accra, Ghana, August 22, 2018; with Prof. Dr. Olivier Menouken Pamen, of AIMS, Ghana}

Now, the issue is to find the right speed of decay of $\varphi(k)$. This is addressed by our theorem. It is important to note that our theorem is useful because not all $\varphi(k)$ work. For example, the seemingly natural choice of $\varphi(k)=\frac{1}{k}$ does not  satisfy the assumptions of our theorem and one can check that it does not have a finite maximizer greater or equal to 2. Other examples of functions that do not satisfy the assumptions of our theorem are $\varphi(k)=1/k^r$ for $r\geq 1$ or, more generally, $\varphi(k)=1/e^{rk}$ for $r\geq 1$

\section{EXAMPLE}
In this section, we provide a function that satisfies our theorem and whose maximizer is close to 5, i.e. $x^*\approx 5$. As explained in \emph{Banker To The Poor} (1998), this is the group size proposed by Muhammad Yunus. Let us consider the following choice $f(x)$, note that it is a function of the size $x$ of the group:
\begin{equation}
    f(x)=x^\alpha + \left(\ln x\right)^{\beta} \label{example}
\end{equation}
This example captures two different forces at play to avoid default. The part of $f(x)$ given by $x^\alpha$ represents the contribution of the material resources available to the group such as the amount of land they possess or the collective financial resources. Meanwhile, the component $\left(\ln x\right)^{\beta}$ represents the contribution of the non material resources of the group; for example, the quality of the group, the peer pressure, or the information available to the group. As the group size increases, there are more collective resources available and thus, the probability of default decreases. We chose $\ln x$, which has a distinctly slower growth rate than $x$, for non material resources of the group because we think that this is less relevant than the material resources.

% \begin{equation}
%     f(x)=x^p +[\ln x]^{\frac{1}{p}} \qquad \text{for all } p\in\left[\frac{1}{2}, 1\right] \label{example}
% \end{equation}
% This example captures two different forces at play. The part of $f(x)$ given by $x^p$ represents the leader's ability to influence the group's performance, while the component $[\ln x]^{\frac{1}{p}}$ represents the quality of the group at play. We chose $\ln x$, which has a distinctly slower growth rate than $x$, for quality of the group because we think that this is less relevant than the leader's ability.

Let us consider $\alpha=p$ and, for simplicity, its reciprocal, i.e., $\beta=1/p$, for all  $p\in\left[\frac{1}{2}, 1\right]$. The exponents of $x$ and of $\ln x$ are chosen in this way because we want to have countervailing forces for the interaction of the material and non-material resources of the group. That is, the less contribution of material resources, the more contribution of non material resources we need. Of course, there is no need to consider $\beta = 1/p$, but, as done in Appendix \ref{Analysis.Example}, this is chosen for mathematical convenience. Other choices that work are $\alpha=p$, $\beta=1/p^2$ or $\beta=1/p^3$ for all $p\in\left[\frac{1}{2}, 1\right]$, among others.

In Appendix \ref{Analysis.Example}, we show that the example given in (\ref{example}) satisfies the conditions of the theorem, but let us now note that the cases $p=\frac{1}{2}$ and $p = 1$ are relevant as calculations show they lead to $x^* = 5.13$ and $x^*=4.62$ respectively. Therefore, in the extreme cases, i.e., a great contribution of either material ($p=1$) or non material collective resources ($p=1/2$), the optimal group size is 5, which coincides with the maximizer proposed by Muhammad Yunus. More generally, we can see that when $p\in\left[0.5, 0.539\right] $ or $p\in\left[ 0.993, 1\right] $, $x^* \in [4.5, 5.5)$, giving an integer maximizer of 5.

We wish to note that, although $f(x) = x^p, \ \text{for all } p \in \left[1/2, 1\right)$ works with our theorem, we believe this function does not capture the complexity of the situation we are trying to model. For this function, when $p$ is close to 1, eg. $p=0.999$, the maximizer is $x^*=503.45$. This should not be surprising as $x^{0.999}$ is close to $x$, which as previously discussed does not have a finite maximizer. Moreover, when $p$ is close to $1/2$, e.g. $p=0.501$, $x^* =1.956 $. A similar, but less drastic situation occurs with $f(x) = (\ln x)^{1/p}, \ \text{for all } p \in \left[1/2, 1\right]$

Other examples of functions that work with our theorem are the following:
\begin{itemize}
    \item $f(x) = \left( x \ln(x) \right)^p$, $p\in[1/2, 3/4]$. By numerical calculations, the maximizer $x^*$ is between 5.17 and 25.52, depending on the value of $p$
    \item $f(x) = \left(\ln( \ln(x) ) \right)^p$, $p>0$ where we necessarily need to consider $x\geq e^e$ to satisfy $f(x)\geq 1$, which is a modified version of condition 1 of our theorem. Then, for example, if $p=0.1$, $x^*=18.23$, if $p=1$, $x^*=22.28$, and if $p=10$, $x^*=309.77$
    \item $f(x)= (x\ln(x))^p+\left(\ln(\ln(x))\right)^{1/p}$, $p\in[1/2, 3/4]$. By numerical calculations, the maximizer $x^*$ is between 6.56 and 18.67, depending on the value of $p$
\end{itemize}

\section{CONCLUSIONS}
In this paper we construct a theoretical model for the determination of the optimal number of people in a group loan. As these loans are intended for low-income borrowers with little or no collateral, and with no credit history, one of the starting points to maximize the repayment rate is to determine the best possible size of the group. We discuss a theorem that provides sufficient conditions for the optimal group size to be finite, greater than 1, and unique. We also provide examples of functions that satisfy our theorem and we analyze in detail the one that we believe has a more natural interpretation and whose associated optimal group size is approximately 5, the number chosen by Muhammad Yunus. An empirical study of the proposed model awaits subsequent research.

\pagebreak

\appendix
\section {APPENDIX A -- Proof of the theorem} \label{Proof.Theorem}

    Let $\mathcal{S}(x):=\sum_{n=0}^\infty \left(\frac{1}{n+1}\right) \left(\frac{1}{n+2}\right) \left( \frac{1}{f(x)} \right)^n $ \par
    Note: 
    \begin{itemize}
        \item $\mathcal{S}(x)$ is a decreasing function in $x$.
        \item $\mathcal{S}(x)\in \left(\frac{1}{2}, 1\right), \ \text{for all } x\in\mathbb{R}$ because:
        $$ \frac{1}{2} < \frac{1}{2} + \sum_{n=1}^\infty \left(\frac{1}{n+1}\right) \left(\frac{1}{n+2}\right) \left( \frac{1}{f(x)} \right)^n = \mathcal{S}(x) < \sum_{n=0}^\infty \left(\frac{1}{n+1}\right) \left(\frac{1}{n+2}\right) = 1 $$
    \end{itemize}
    
    Set $h(x):=f(x)-xf'(x)$ and note that in $(a,b)$, $h(x)$ is a monotone function because of condition (\ref{itm:4ii} or (\ref{itm:4iv}. More explicitly:
    $$ h'(x)=f'(x)-\left[ f'(x)+xf''(x) \right] = -xf''(x) >0 \ (\text{or } <0) \qquad \text{for all } x\in(a, b) $$ 
    Moreover, the monotonicity of $h(x)$ and condition (\ref{itm:4i} or (\ref{itm:4iii} imply $\frac{1}{2}<h(x)<1$ \ $\text{for all } x \in (a,b)$
    
    In this way $\text{for all } x\in(a,b)$:
    \begin{itemize}
        \item Both $\mathcal{S}(x)$ and $h(x)$ are continuous and monotone
        \item $\mathcal{S}(x)$ is bounded between $\left( \frac{1}{2}, 1 \right).$ This actually holds $\text{for all } x\in\mathbb{R}^+$ 
        \item $h(x)$ increases from $\frac{1}{2}$ to $1$ (or decreases from $1$ to $\frac{1}{2}$)
    \end{itemize}
    Then there exists a unique $x^*\in(a,b)$ such that 
    \begin{equation} \label{eq.proof}
        h(x^*)=\mathcal{S}(x^*) 
    \end{equation} 
    
    We shall see that this $x^*$ is actually the unique maximizer. Thanks to equation (\ref{eq.proof}), we have:
    \begin{align}
        f(x^*) - x^* f'(x^*) & =\sum_{n=0}^\infty \left(\frac{1}{n+1}\right) \left(\frac{1}{n+2}\right) \left( \frac{1}{f(x^*)} \right)^n \nonumber \\
        \iff 0 &= \sum_{n=0}^\infty \left(\frac{1}{n+1}\right) \left(\frac{1}{n+2}\right) \left( \frac{1}{f(x^*)} \right)^n +x^*f'(x^*)-f(x^*) \nonumber \\
        &= \sum_{n=0}^\infty \left(\frac{1}{n+1}\right) \left( \frac{1}{f(x^*)} \right)^n - \sum_{n=0}^\infty \left(\frac{1}{n+2}\right) \left( \frac{1}{f(x^*)} \right)^n +x^*f'(x^*)-f(x^*) \nonumber \\
        &= \sum_{n=1}^\infty \left(\frac{1}{n}\right) \left( \frac{1}{f(x^*)} \right)^{n-1} - \sum_{n=2}^\infty \left(\frac{1}{n}\right) \left( \frac{1}{f(x^*)} \right)^{n-2} - f(x^*) +x^*f'(x^*) \nonumber \\
        &= \sum_{n=1}^\infty \left(\frac{1}{n}\right) \left( \frac{1}{f(x^*)} \right)^{n-1} - \sum_{n=1}^\infty \left(\frac{1}{n}\right) \left( \frac{1}{f(x^*)} \right)^{n-2}  +x^*f'(x^*)  \nonumber \\
        &= \sum_{n=1}^\infty \left(\frac{1}{n}\right) \left( \frac{1}{f(x^*)} \right)^{n+1} - \sum_{n=1}^\infty \left(\frac{1}{n}\right) \left( \frac{1}{f(x^*)} \right)^{n}  + \frac{x^*f'(x^*)}{(f(x^*))^2}  \label{eq.proof.iff}
    \end{align}
    
    Now, recall we want to find a maxima for $\left(1-\varphi(x)\right)^x$. This is equivalent to maximizing $\mathcal{U}(x):= x\ln\left(1-\varphi(x)\right)$.\par
    Note $\mathcal{U}'(x)= \ln\left( 1- \varphi(x) \right) - \frac{x}{1-\varphi(x)} \varphi'(x)$ \par
    It suffices to find $x^*$ (the maximizer) such that $\mathcal{U}'(x^*)=0$, which is equivalent to $ g(x^*)=0 $
    where $g(x):= \left[ 1-\varphi(x) \right] \ln\left(1-\varphi(x)\right) - x\varphi'(x)$\par
    Recall: $\ln(1-y) = - \sum_{n=1}^\infty \frac{y^n}{n}$, if $|y|<1$. Then:
    \begin{align*}
        g(x) & = \left[ 1-\varphi(x)\right] \left[ -\sum_{n=1}^\infty \frac{\varphi^n(x)}{n} \right] - x\varphi'(x) \\
        & =\sum_{n=1}^\infty \frac{\varphi^{n+1}(x)}{n} -\sum_{n=1}^\infty \frac{\varphi^n(x)}{n} -x\varphi'(x) \\
        & =\sum_{n=1}^\infty \left( \frac{1}{n}\right) \left(\frac{1}{f(x)} \right)^{n+1} - \sum_{n=1}^\infty \left( \frac{1}{n}\right) \left(\frac{1}{f(x)} \right)^{n} + x \frac{f'(x)}{\left( f(x) \right)^2}
    \end{align*}
    Finally, it is easy to see that this last line and (\ref{eq.proof.iff}) imply $g(x^*) =0$ and we can conclude $x^*$ is the unique maximizer in $(a,b)$. If $a, b$ are unique, it is clear that the maximizer is unique.
    
\section{APPENDIX B -- Analysis of the example} \label{Analysis.Example}
Now, we show that the example given in (\ref{example}) satisfies the conditions of the theorem.

Conditions 1 and 2 are immediate. \par

\textbf{Condition (3):} $f'(x)>0$ $\text{for all } x\geq 2$ \par
\textbf{Proof of (3):} $f'(x)=px^{p-1} + \frac{1}{p} \left(\frac{1}{x}\right) \left( \ln x \right)^{\frac{1}{p}-1} $ \qquad As $x\geq 2$, it is clear $f'(x)>0$ \quad \qedsymbol \par 

\vspace{3mm}

\textbf{Condition (\ref{itm:4i}:} There exist $a$ and $b$ such that $f(a)-af'(a)=\frac{1}{2}$ and $f(b)-bf'(b)=1$ \par

\textbf{Proof of (\ref{itm:4i}:} \par 
Set $h_p(x):=f(x)-xf'(x)= x^p + (\ln x)^{\frac{1}{p}} - px^p -\frac{1}{p} \left(\ln x\right )^{\frac{1}{p}-1} = (1-p)x^p + \left(\ln x\right)^{\frac{1}{p}-1} \left(\ln x -\frac{1}{p} \right) $ \par 

\textbf{Claim:} $h_p(x)$ is increasing in $x$\par
\textbf{Proof of claim:} \par
$\frac{\partial}{\partial x} h_p(x) =(1-p)px^{p-1} +\left( \frac{1}{p} -1 \right) \left( \ln x \right)^{\frac{1}{p}-2} \left[ \ln x -\frac{1}{p} \right]  \frac{1}{x}  + \frac{1}{x} \left( \ln x \right)^{\frac{1}{p}-1} $ \par
It is clear $ \left( 1- p \right) p x^{p-1} > 0 $. So, it suffices to show 
\begin{equation}\label{eq:Proof-example}
    \left( \frac{1}{p} -1 \right) \left( \ln x \right)^{\frac{1}{p}-2} \left[ \ln x -\frac{1}{p} \right]  \frac{1}{x}  + \frac{1}{x} \left( \ln x \right)^{\frac{1}{p}-1} \geq  0
\end{equation}
For reasons that will become clear later, we only consider $x\geq e$. As $\ln x +1 \geq 2$ and $ 1 \leq \frac{1}{p} \leq 2$, it follows that 
\begin{align*}
    \ln x +1 & \geq  \frac{1}{p} \\
    \frac{1}{p} \left(\ln x  +1 - \frac{1}{p} \right) & \geq 0  \\
    \frac{1}{p} \left( \ln x - \frac{1}{p} \right) - \left( \ln x - \frac{1}{p} \right) + \ln x & \geq 0 \\
     \left( \frac{1}{p} -1 \right)  \left( \ln x -\frac{1}{p} \right) +   \ln x & \geq 0 
\end{align*}
This shows \eqref{eq:Proof-example} and thus that $h_p(x)$ is increasing for $x \geq e$ \quad \qedsymbol \par

\textbf{Claim:} $h_p(e)=(1-p) e^p + 1 - \frac{1}{p}$ is concave in $p$ and hence there exists a local maxima, namely $p^*$. (Recall $\frac{1}{2} \leq p \leq 1$)\par
\textbf{Proof of claim:} \par
$\frac{\partial}{\partial p} h_p(e)= -e^p + (1-p) e^p + \frac{1}{p^2} = -pe^p + \frac{1}{p^2}$ \par 
$\frac{\partial^2}{\partial p^2} h_p(e) = -e^p -pe^p -  \frac{2}{p^3} < 0 \Longrightarrow \ \ h_p(e) $ is concave \par 
Now, to find the maxima, we set the derivative equal to 0, i.e. $\frac{\partial}{\partial p} h_p(e) = 0$ 
\begin{align*}
    0 & = -pe^p + \frac{1}{p^2} \\
    1 & = p^3 e^p \\
    1 & = p e^{\frac{1}{3}p} 
\end{align*}
Set $u = \frac{1}{3} p$, we need to solve $u e^u = \frac{1}{3}$, which we do by using the product logarithm. \par
Hence $u=W\left( \frac{1}{3} \right)$ and thus $p^*=3W\left( \frac{1}{3} \right) \approx 0.772883 \Longrightarrow h_p(e)|_{p=0.773} \approx 0.1981 < \frac{1}{2}$ \quad \qedsymbol 

\textbf{Claim:} $h_p(e) < \frac{1}{2} \ \ \text{for all } p \in \left[ \frac{1}{2}, 1 \right]$  \par

\textbf{Proof of claim:} \par
As we have shown that $h_p(e)$ is concave in $p$ and that $h_{p^*}(e) \approx 0.1981 < \frac{1}{2}$, it follows that $h_p(e) < \frac{1}{2} \ \ \text{for all } p \in \left( \frac{1}{2}, 1 \right)$ \par 
We only need to check the endpoints, $h_p(e)|_{p=\frac{1}{2}} \approx -0.1756 $ and $h_p(e)|_{p=1}=0$ \quad \qedsymbol \par
Hence $h_p(e) < \frac{1}{2} \ \  \text{for all } p\in\left[ \frac{1}{2}, 1 \right] $. This, along with $h_p(x)$ being continuous, increasing in $x\geq e$ and $\lim_{x\rightarrow \infty} h_p(x)= \infty$, imply that there exists $ a $ such that $h_p(a) = \frac{1}{2}$ and that $a \geq e \approx 2.7 \Rightarrow a \geq 3$ \par 

\vspace{5mm}
\textbf{Claim:} $h_p(e^2) = (1-p) e^{2p} + 2^{\frac{1}{p}-1} \left( 1-\frac{1}{p} \right)$ is concave in $p$ \par
\textbf{Proof of claim:} \par
$\frac{\partial}{\partial p} h_p(e^2) = -e^{2p} + 2(1-p)e^{2p} +\left( \frac{1}{p^3} -\frac{2}{p^2} \right) (\ln 2) 2^{\frac{1}{p}-1} + \left(\frac{1}{p^2}\right) 2^{\frac{1}{p}-1} $ \par 
$\frac{\partial^2}{\partial p^2} h_p(e^2) = -4e^{2p} + 4(1-p)e^{2p} - 2^{\frac{1}{p}} \frac{1}{p^3} + 2^{\frac{1}{p}} \left( 2 -\frac{1}{p} \right) \frac{\ln 2}{p^3} - 2^{\frac{1}{p}} \frac{\ln 2}{p^4} + 2^{\frac{1}{p}-1} \left( 2-\frac{1}{p} \right) \frac{(\ln 2)^2}{p^4}$ \par 
Now we show that $\frac{\partial^2}{\partial p^2} h_p(e^2)<0$  
\begin{enumerate}
    \item It is clear that $-4e^{2p} + 4(1-p)e^{2p} <0$
    \item As $\ln 2 <1$ and $2-\frac{1}{p} \leq 1$
        \begin{align*}
            \left(2-\frac{1}{p}\right) \ln 2 < & \ 1\\
            -1 +\left(2-\frac{1}{p}\right) \ln 2 < & \ 0 \\
            - 2^{\frac{1}{p}} \frac{1}{p^3} + 2^{\frac{1}{p}} \left( 2 -\frac{1}{p} \right) \frac{\ln 2}{p^3} < & \ 0
        \end{align*}
    \item Similarly
        \begin{align*}
            \left(2-\frac{1}{p}\right)\frac{\ln2}{2} < & \ 1 \\
            -1 +\left(2-\frac{1}{p}\right)\frac{\ln2}{2}< & \ 0 \\
            2^{\frac{1}{p}} \frac{\ln 2}{p^4} + 2^{\frac{1}{p}-1} \left( 2-\frac{1}{p} \right) \frac{(\ln 2)^2}{p^4}< & \ 0 \qquad \qedsymbol \\
        \end{align*} 
\end{enumerate}
\textbf{Claim:} $h_p(e^2) \geq 1 \ \ \text{for all } p\in\left[ \frac{1}{2}, 1\right]$ \par 
\textbf{Proof of claim:} \par 
As we have shown that $h_p(e^2)$ is concave in $p$, it suffices to show $h_p(e^2)|_{p=\frac{1}{2}}\geq 1$ and $h_p(e^2)|_{p=1}\geq 1$ \par
\begin{enumerate}
    \item $h_p(e^2)|_{p=\frac{1}{2}} =\frac{1}{2} e > 1$ as $e>2$
    \item $h_p(e^2)|_{p=1} =1$ \quad \qedsymbol
\end{enumerate}
Hence $h_p(e^2) \geq 1 \ \  \text{for all } p\in\left[ \frac{1}{2}, 1 \right] $. This, along with $h_p(x)$ being continuous and increasing in $x$ implies that there exists $ b $ such that $h_p(b) = 1$ and that $b \leq e^2 \approx 7.4 \Rightarrow b \leq 7$ \quad \qedsymbol \par 

\vspace{5mm}

\textbf{Condition (\ref{itm:4ii}:} $f''(x)<0$ $\text{for all } x\in(a,b)$ for some $a, b>0$ such that $e\leq a<b$ \par 

\textbf{Proof of (\ref{itm:4ii}:} \par 
$f''(x)=p(p-1)x^{p-2} +\frac{1}{p}\left[ \left( \frac{1-p}{p} \right) \left( \frac{1}{x} \right)^2 \left( \ln x \right)^{\frac{1}{p}-2} - \left(\frac{1}{x}\right)^2 \left( \ln x \right)^{\frac{1}{p}-1} \right] $ \par
\ \ \ \ \ \quad $= p(p-1) x^{p-2} + \frac{1}{p} \left( \frac{1}{x} \right)^2 \left( \ln x \right)^{\frac{1}{p}-2} \left[ \frac{1-p}{p} -\ln x \right]$ 
\begin{enumerate}
    \item $p(p-1)x^{p-2} \leq 0$ because, by assumption $p\leq 1$. \par 
    \item By assumption $1\leq \frac{1}{p} \leq 2$ and as $\ln x +1 \geq 2 $ for $x\geq e$, we get $\frac{1}{p} \leq  \ln x +1$ . Hence, $\frac{1-p}{p} - \ln x \leq 0$, which implies the 2nd term is non-positive for all $x\geq e$ \quad \qedsymbol
\end{enumerate}

Hence for $f(x)=x^p +[\ln x]^{\frac{1}{p}}$, using our theorem, we can claim that the maximizer $x^*\in[e, e^2] \ \text{for all } p\in\left[\frac{1}{2}, 1\right]$. It is worth noticing that for this particular $f(x)$, we can obtain a narrower interval in the following way:

To find the maximizer $x^*$ of $\left( 1-\varphi(x) \right)^x$, we need to set the derivative equal to $0$, which, as noted in the proof of the theorem, is equivalent to solving $\left[ 1-\varphi(x) \right] \ln\left(1-\varphi(x)\right) - x\varphi'(x) = 0 $. Using $\varphi(x) = \frac{1}{f(x)}= \frac{1}{x^p +[\ln x]^{1/p}}$, let us define:
\begin{align*} 
    \mathcal{H}(x,p):& =  \left[ 1-\varphi(x) \right] \ln\left(1-\varphi(x)\right) - x\varphi'(x) \\
    & =  \left( 1 - \frac{1}{x^p +[\ln x]^{\frac{1}{p}}} \right) \ln \left( 1 - \frac{1}{x^p +[\ln x]^{\frac{1}{p}}} \right) +  \frac{p^2 x^p \ln x + \left( \ln x \right)^{\frac{1}{p}}}{p \ln x \left[ x^p + \left( \ln x \right)^{\frac{1}{p}} \right]^2} 
\end{align*}

Then, after fixing $p$, we need to find $x$ such that $\mathcal{H}(x, p) = 0$. As our theorem allows us to conclude that the maximizer $x^*\in[e, e^2]$, we only need to analyze the behaviour of $\mathcal{H}(x,p)$ when $x$ is in such interval.

Note $\mathcal{H}(x, p)$ is decreasing in $x$ on the interval $x\in[e, e^2]$ for all fixed   $p \in \left[ \frac{1}{2}, 1 \right]$. Now, we want to find a value of $x\in[e, e^2]$ for which $\mathcal{H}(x, p) \geq 0$ for all $p \in \left[ \frac{1}{2}, 1 \right]$. Since this is not the case for $x=3.5$, we choose $x=3.4$ as such bound gives us uniform positivity for all $p \in \left[ \frac{1}{2}, 1 \right]$

More precisely, for fixed $p \in \left[\frac{1}{2}, 1\right] $ and $\text{for all } x \leq 3.4 $, we have $\mathcal{H}(x,p) \geq \mathcal{H}(3.4,p)$, which implies
\begin{equation*}
   \mathcal{H}(x,p) \geq \min_{p} \mathcal{H}(x,p) \geq  \min_{p}\mathcal{H}(3.4,p) > 0
\end{equation*}
Hence $\mathcal{H}(x,p) = 0$ does not have a solution when $x \leq 3.4 $ and $ p \in \left[ \frac{1}{2}, 1\right]$. So, ${x^*}$  must be in the interval $(3.4, \infty)$.

Similarly, we want to find a value of $x\in[3.4, e^2]$ for which $\mathcal{H}(x, p) \leq 0$ for all $p \in \left[ \frac{1}{2}, 1 \right]$. We can check that $x=5.2$ gives us uniform negativity for all $p \in \left[ \frac{1}{2}, 1 \right]$. In this way we get $\text{for all } x\geq 5.2$ and $\text{for all } p \in \left[ \frac{1}{2}, 1\right] $:
$$ \mathcal{H}(x,p) \leq \max_p\mathcal{H}(x, p) \leq \max_p \mathcal{H}(5.2, p) < 0 $$
This implies that $\mathcal{H}(x,p) = 0$ does not have a solution when $x \geq 5.2 $ and $ p \in \left[ \frac{1}{2}, 1\right]$. Hence, we finally get a narrower interval $x^* \in (3.4, 5.2)$ $ \text{for all } p\in \left[ \frac{1}{2}, 1\right]$ 

Note that the choice of $x=3.4$ and $x=5.2$ as a comparison points was arbitrary as all we require is $ \mathcal{H}(x,p)$ to be positive or negative (respectively) for all $p\in \left[ \frac{1}{2}, 1\right]$. Another options that work are any $e\leq x\leq 3.486$ and $5.135 \leq x \leq e^2$ respectively, but as we will round up or down to the nearest integer, we believe that using one decimal place is enough.

\pagebreak

% \printbibliography[title={REFERENCES}, heading=bibintoc] 
\bibliographystyle{authordate1}
\bibliography{main.bib} 

\end{document}